\documentclass{aa}

\usepackage{newtxtext,newtxmath}
\usepackage[T1]{fontenc}
\usepackage[dvipsnames]{xcolor}
\usepackage{wasysym}
\usepackage{graphicx}
\usepackage{hyperref}  
\usepackage{xspace}
\usepackage{caption}
\usepackage{placeins}
\usepackage{xcolor}
\usepackage{subcaption}

\begin{document}
\title{Influence of mergers on LyC escape of high redshift galaxies}

\author{Ivan Kostyuk\inst{1, 2}, Benedetta Ciardi\inst{1}}

\institute{Max-Planck-Institut f\"{u}r Astrophysik, Karl-Schwarzschild-Str. 1, 85741 Garching, Germany \\ \and Scuola Normale Superiore, Piazza dei Cavalieri 7, I- 50126 Pisa, Italy \\ {E-mail: ivan.kostyuk@sns.it}}


\abstract
{}
{We investigate the impact of galaxy mergers on the Lyman Continuum (LyC) radiation escape, $f_{\rm esc}$, from high-redshift galaxies.}
{We post-process $\approx 5 \times 10^5$ galaxies (redshift $5.2 < z < 10$) extracted from the TNG50 cosmological simulation using a physically motivated analytic model for LyC escape.}
{Galaxies that have not experienced a merger for the last $\approx 700$~Myr have an average  $f_{\rm esc} \sim 3\%$, which increases to up to 14\% immediately following a merger. The strongest effect can be observed in galaxies with stellar masses of $\sim 10^7$~M$_\odot$. 
We attribute the increase in the escape fraction to two main factors: ({\it i}) accretion of metal-poor gas onto the central region of a galaxy, which feeds star formation and LyC emission; and ({\it ii}) displacement of neutral gas relative to star-forming regions, which reduces the optical depth to LyC photons. 
We additionally examine how proximity to other galaxies influences LyC escape, finding that galaxies with more neighbors tend to have more frequent mergers, and thus a higher LyC leakage. However, galaxies in overdense regions tend to have a larger LyC escape fraction independently from mergers, because of their higher gas inflow, and consequent increase in the star formation rate. 
The increase in both mergers and gas inflow could contribute to low-mass galaxies ionizing proximity zones of high-$z$ Ly$\alpha$ leakers recently observed with JWST.}
{}


\maketitle
\section{Introduction}
\label{sec:intro}

The reionization of the intergalactic medium (IGM) is the last major phase transition undergone by our Universe. Although a consensus has been reached within the community that the process is mainly driven by stellar type sources, more energetic sources such as accreting black holes or X-ray binaries are expected to contribute to the heating and reionization of the IGM (see e.g. \citealt{Qin2017,dayal2020,eide20}).
Constraints on cosmic reionization are limited to an estimate of the global amount of electrons produced during the process through measurements of the Thomson scattering optical depth \citep{planck2020}, and of the amount of neutral hydrogen in the IGM during the latest stages of reionization mainly through observations of Ly$\alpha$ emitting galaxies and spectra of quasars and gamma ray bursts (e.g. \citealt{totani2014,banados2018,ouchi2020,bosman2022}). 
Observations of the high redshift universe have been revolutionized by the advent of the James Webb Space Telescope\footnote{https://webb.nasa.gov/} (JWST), which is providing increasingly richer information on the properties of high redshift sources (e.g. \citealt{ArrabalHaro2023,Bouwens2023,donnan2023}). 

A key ingredient for the understanding of both reionization process and galactic properties is the escape fraction of ionizing photons, a quantity which remains highly unconstrained by observations (e.g. \citealt{vanzella2018,bian20,bosman20}), and at the same time is extremely difficult to model due to the wide range of scales involved, as well as the dependence on a large number of galactic properties (e.g. \citealt{paardekooper15,rosdahl18,ma20,ocvirk21,smith22,kostyuk23a}).

As direct observations of Lyman continuum (LyC) radiation escaping from high redshift galaxies is hampered by the large IGM opacity, observational efforts have concentrated at intermediate ($z \sim$ 3) and low ($z \sim$0.2) redshift, and have resulted in a steady increase in the number of known LyC leakers (e.g. \citealt{flury2022}). A recent key development has been the identification of possible indirect diagnostics of LyC escape, such as the slope of the LyC radiation (e.g. \citealt{chisholm2022}), the prominence of certain metal lines (e.g. \citealt{barrow20}, \citealt{nakajima20}, \citealt{katz20}), or the details of the Ly$\alpha$ emission line (e.g. \citealt{dijkstra14}, \citealt{chisholm18}).

Recently, measurements of the HI 21cm emission in the nearby reionization-era analog galaxy Haro 11 presented by \cite{lereste2023} have identified galactic mergers as potential facilitators of LyC production and escape in high redshift galaxies. In particular, the study suggests that repeated gas compression during a merger could drive star formation, thereby increasing LyC production and feedback that clears out channels, facilitating LyC escape. Furthermore, the process of tidal gas displacement removes some of the gas around star-forming regions, decreasing LyC photon absorption. In line with these findings, \cite{witten23} have proposed that the increased frequency of merger events in overdense regions at high redshifts ($z$ > 7) might contribute to the formation of the large ionized bubbles necessary for the detection of Ly$\alpha$ emitters at these redshifts.

The goal of this work is to use simulations to investigate the physical impact mergers have on early dwarf galaxies that affect LyC escape. To this end, we extend the semianalytic framework developed by \cite{kostyuk23b} (hereafter K24) based on the physical model developed by Ferrara et al. (2025; hereafter F25) to describe the dependence of the escape fraction on a number of galactic properties following galactic mergers. The aim hereby is to qualitatively understand how changes in physical properties following a merger impact the LyC escape of galaxies. This paper is structured as follows. In Section~\ref{sec:method} we provide a brief introduction to the methodology used for our analysis, in Section~\ref{sec:res} we discuss the results, and in Section~\ref{sec:concl} we provide our conclusions.

\section{Methodology}
\label{sec:method}
In this Section we introduce the basics of the methodology used for our analysis, while we refer the reader to K24 for more details.
\subsection{Illustris TNG50 simulation}
\label{sec:tng}
Illustris TNG50 \citep{nelson19a,pillepich19} is a large scale cosmological simulation of length $51.7\mathrm{cMpc}$, with initial conditions generated using the N-GenIC code \citep{springel05} and cosmological parameters   $\Omega_\mathrm{m}=0.3089$, $\Omega_\mathrm{b}=0.0486$, $\Omega_\Lambda=0.6911$, $h=0.6774$, $\sigma_8=0.8159$ and $n_s=0.9667$ \citep{planck16}. 
The simulation is performed on a Voronoi grid using the code {\textsc AREPO} \citep{springel10}.
Following \cite{weinberger17} and \cite{pillepich18a}, the gas particles contained within Voronoi cells with hydrogen number density larger than $0.1 \mathrm{cm}^{-3}$ are stochastically converted into stellar particles, each representing a stellar population with a Chabrier initial mass function (IMF; \citealt{chabrier03}). 
The \cite{springel03} model for the interstellar medium (ISM) is adopted to include sub-grid processes and pressure support.
The initial gas and dark matter particles mass is $8.5 \times 10^4 \mathrm{M}_\odot$ and $4.5 \times 10^5\mathrm{M}_\odot$, respectively, but these values can change up to a factor of two due to the refinement and de-refinement of the cells.
As a reference, the mean size of star forming gas cells is 144~ppc at $z=6$.

Using the SubFind \citep{springel01} and friends-of-friends  \citep{davis85} algorithms, galaxies are identified within sub-halos of minimum mass $\sim 10^6\mathrm{M}_\odot$. A black hole of mass $8\times 10^5 h^{-1} \mathrm{M}_\odot$ is seeded in galaxies which reach a mass of  $10^{10.8}\mathrm{M}_\odot$, and it subsequently grows through mergers and gas accretion, according to an Eddington limited Bondi formalism. Feedback from black holes and supernovae is modeled as in \cite{weinberger17} and \cite{pillepich18a}, respectively. Metal enrichment is included following the individual abundances of elements such as C, N, O, Ne, Mg, Si, Fe, and Eu. Finally, a spatially uniform UV-background (from \citealt{fg09}) is turned on at $z<6$. 

We refer the reader to \cite{nelson19a} and \cite{pillepich19} for further details.

\subsection{Escape fraction model}
\label{sec:escfra}
In this section we briefly describe the method used to model the escape fraction of ionizing photons, while we refer the reader to F25 and K24 for more details.

In this approach a galaxy is modeled as a plane parallel slab\footnote{See K24 on a discussion of the validity of this model for non disc-like galaxies.}, with ionized radiation produced within the galactic plane, while the gas is distributed above and below it with constant density, $n$, up to a gas scale height $H$. This simplified model is a reasonable approximation given the fact that most of the gas is concentrated within only a few cells in proximity to the galactic disc. Ionizing photons are absorbed by gas as well as dust, and are able to escape from the galaxy only if their flux, $F_i$, is strong enough to form a channel of ionized gas. By expressing the ionizing energy flux as a power-law with slope $\beta=4$, and using the conversion between UV luminosity and star formation surface density, $\Sigma_\mathrm{SFR}$, adopted by the REBELS collaboration \citep{dayal2020}, we obtain (see eqs.~1-3 in K24):
\begin{equation}
    F_i = \frac{\Sigma_\mathrm{SFR}}{\beta h_P \kappa}, 
    \label{eq:flux}
\end{equation}
with $h_P$  Planck constant, and $\kappa =7.1\times 10^{-29}$ $\mathrm{M}_\odot$ $\mathrm{yr}^{-1}$$\mathrm{erg}^{-1}$ $\mathrm{s}\mathrm{Hz}$ a conversion factor appropriate for the Chabrier IMF used in TNG50.

The ionizing radiation emitted by the stars is absorbed both by neutral hydrogen (HI) and dust, with optical depth:
\begin{equation}
    \tau_\mathrm{HI} = \frac{N_0}{N_S} 
    \;\;\; {\rm and} \;\;\;\tau_\mathrm{d} = \frac{N_0}{N_d}.   
\end{equation}
Here, $N_0=n H$ is the gas column density, $N_S=F_i/(n \alpha_{\rm B})$ is the maximum gas column density that can be ionized by the flux $F_i$, $\alpha_B$ is the case B recombination rate, $N_d = 4.3 \times 10^{20} \mathcal{D}^{-1} \mathrm{cm}^{-2}$ is the gas column density at which the dust optical depth reaches unity, and $\mathcal{D}= Z/Z_\odot$ is the dust-to-gas ratio normalized to the galactic value obtained assuming a linear relation between metallicity $Z$ and dust (see eqs. 4-7 in K24).

We then write the LyC escape fraction as $f_\mathrm{esc} = e^{-(\tau_\mathrm{HI}+\tau_{d})}$, i.e.:
\begin{equation}
    f_\mathrm{esc} = \exp \left[-N_0\left(\frac{1}{N_S}+ \frac{1}{N_d} \right) \right]. \label{eq:esc}
\end{equation}

To evaluate the escape fraction of a galaxy, we define the galactic plane as the plane spanned by the first two principle axes of the galaxy, and from it we extract a cube of length equivalent to 4 times the stellar half mass radius. The gas particles contained in the cube are then distributed on a Cartesian grid with $100^3$ cells\footnote{The analysis performed in the Appendix of K24 shows that an excellent convergence in the results is reached with this grid size.}. 
In this framework, $H$ is defined as the height below and above the galactic plane that contains 63\% of the gas mass. All the physical quantities necessary to the calculation of the escape fraction are then projected onto the galactic plane and evaluated in each of the $100^2$ 2D grid cells. The escape fraction of a galaxy is then given by:
\begin{equation}
    f_\mathrm{esc} = \frac{\sum\limits_\mathrm{cell} F_{i,\mathrm{cell}}f_\mathrm{esc,cell}}{\sum\limits_\mathrm{cell} F_{i,\mathrm{cell}}},
\end{equation}
where $F_{i,\mathrm{cell}}$ and  $f_\mathrm{esc,cell}$ are the ionizing flux and escape fraction in a cell obtained using eqs.~\ref{eq:flux} and~\ref{eq:esc} with the value of the physical quantities evaluated in each element of the 2D grid as detailed in K24.

Here, we calculate the escape fraction of $\approx 600,000$ galaxies with a minimum stellar and gas mass of $10^{5.5}\mathrm{M}_\odot$ and $10^{6}\mathrm{M}_\odot$, respectively, from 17 TNG50 snapshots in the redshift range $5.2 \leq z \leq 20$. The time since the last mergers was obtained using catalogues based on the work of \cite{rodrigueq17} and \cite{eisert23}, where the evolution of a galaxy is traced by identifying its particles in subsequent snapshots.

We note that the model presented here is first and foremost designed to qualitatively capture the characteristics of LyC escape and their dependence on galactic properties, rather than to obtain the exact quantitative value of the escape fraction. Indeed, given the highly non-linear nature of LyC escape, this is unlikely to be captured by a semianalytic model. Additionally, as the TNG50 simulation does not fully resolve the interstellar medium (ISM), even with a fully numerical approach as the one presented in \cite{kostyuk23a}, the exact value of the escape fraction depends on the assumptions adopted to model the unresolved scales. We highlight, though, that this problem afflicts all numerical investigations of LyC escape, as none is presently able to resolve all the scales involved, down to the molecular clouds ones.


\section{Results}
\label{sec:res}
In this Section we present the results of our analysis in terms of the global evolution of $f_\mathrm{esc}$, its correlation to various galactic properties and the impact that mergers have on it.

\subsection{Global evolution of $f_\mathrm{esc}$ after mergers}

\begin{figure}
    \centering
    \includegraphics[width=0.49\textwidth]{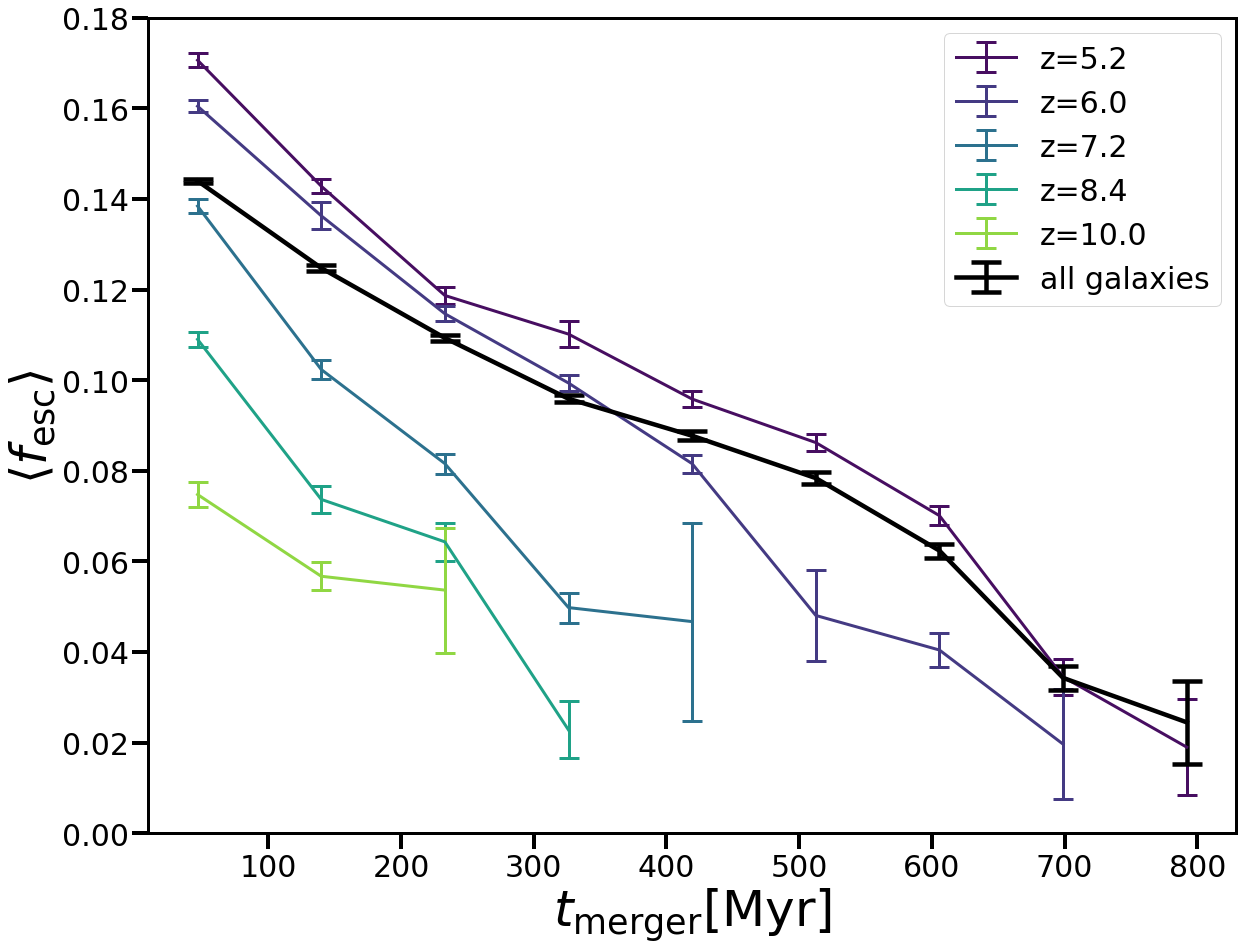}
    \caption{Average LyC escape fraction as a function of time elapsed since the last major merger for galaxies at various redshifts, as indicated by the colors. The black line refers to all redshift combined.}
    \label{fig:redshift}
\end{figure}

In fig.~\ref{fig:redshift}, we begin our investigation by exploring the evolution of the average (including all galaxies) LyC escape $\langle f_\mathrm{esc} \rangle$ as a function of the time elapsed since the most recent major merger (i.e. with a stellar mass ratio $>1/4$ between the two merging galaxies) event, $t_\mathrm{merger}$, at various redshifts.
Here, we see a clear decline of $\langle f_\mathrm{esc} \rangle$ with increasing  $t_\mathrm{merger}$. Although this trend is consistently observed at all redshifts, it is less pronounced at higher redshift, due to the overall lower value of the LyC escape fraction, as previously shown in K24. When all redshifts are analysed together, $\langle f_\mathrm{esc} \rangle$ decreases from $\sim 14\%$ for galaxies which have experienced a merger within the last $\sim 50$~Myr, to $\sim 3\%$ in the case of mergers that took place $\sim 800$~Myr prior. It should be noted that, since more massive galaxies experience fewer mergers, the decrease in $\langle f_\mathrm{esc} \rangle$ with increasing merger time reflects to some extent the fact that typically the LyC escape is lower in more massive galaxies. 

\begin{figure}
    \centering
    \includegraphics[width=0.49\textwidth]{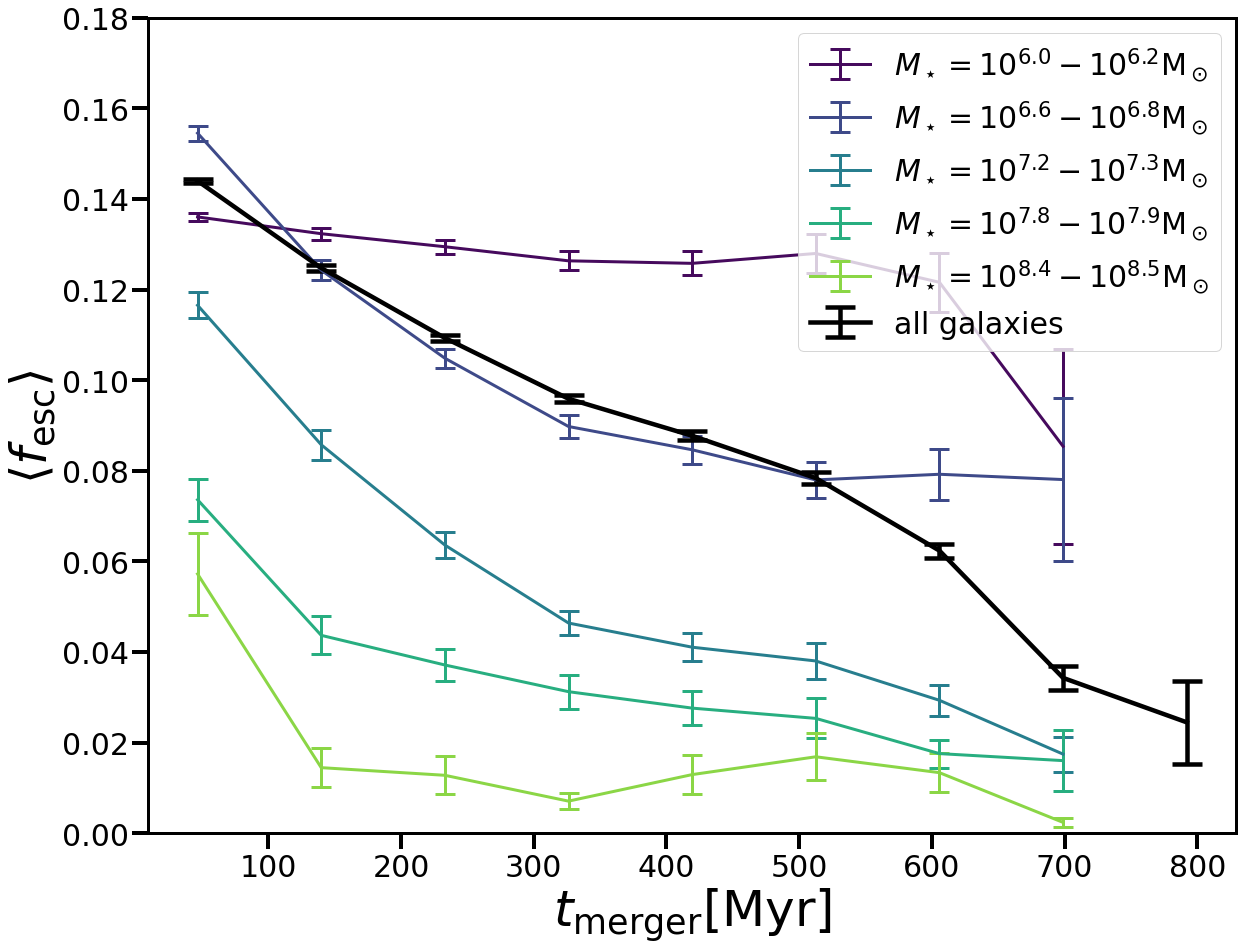}
    \caption{Average LyC escape fraction as a function of time elapsed since the
last major merger for galaxies at all examined redshifts with different stellar masses, as indicated by the colors. The black line refers to all masses combined and is the same as in fig.~\ref{fig:redshift}.
}
    \label{fig:mass}
\end{figure}
Therefore, to disentangle the effects of merger time and galactic mass, in fig.~\ref{fig:mass} we examine the average escape fraction as a function of $t_\mathrm{merger}$ in different stellar mass bins. 
We again observe a decline in $\langle f_\mathrm{esc} \rangle$ with increasing $t_\mathrm{merger}$, but now it is less pronounced, with the average escape fraction decreasing by approximately $2\%$ in the lowest stellar mass bin, i.e. $M_\star = (10^6-10^{6.2})$~M$_\odot$. 
The strongest decline ($\sim 10\%$) is found at $M_\star = (10^{7.2}-10^{7.3})$~M$_\odot$, while for larger masses it becomes smaller again due to the generally lower LyC escape fractions exhibited by more massive galaxies.
The reason for the weaker correlation observed in the lowest mass bin is twofold. First, a significant fraction of galaxies leak LyC through an extended (rather than localized) area which, as will be shown in fig.~\ref{fig:differences_localized_extended}, is affected differently by mergers. Second, the merger events in these galaxies take place on much shorter time scales, such that the changes in galactic morphology and SFR are not resolved with the time resolution used here. 
The suggested shorter time scales of merger induced LyC escape aligns with the patterns identified in prior galactic merger simulations. 
\cite{renaud14} find a sequence of multiple star-formation bursts over a time span of $\sim 100$~Myr subsequent to the merger event involving galaxies with $M_\star\sim  10^{10}\mathrm{M}_\odot$.
Conversely, simulations by \cite{lahen19} focusing on less massive galaxies with $M_\star\sim 10^7\mathrm{M}_\odot$ produce a single burst of star formation as a consequence of the merger, typically lasting only a few Myr. By extrapolating this trend towards lower masses, it can be inferred that galaxies with $M_\star\sim 10^6\mathrm{M}_\odot$ would likely experience even briefer episodes of star formation subsequent to merger occurrences.

We note here that the same investigation has been performed also on the galaxies analysed in \cite{kostyuk23a} with a full radiative transfer approach for the evaluation of the escape fraction. While a similar correlation between $\langle f_\mathrm{esc} \rangle$ and $t_\mathrm{merger}$ is found, it is not shown here as that study was limited to a smaller stellar mass range of $(10^7-10^8)$~M$_\odot$ and to a lower number of galaxies, rendering the results less statistically significant.

\subsection{Impact of galactic mergers on different escape modes}

In K24 we have shown that LyC escapes from early galaxies in two distinct ways. If cooling is efficient in small galaxies ($M_\star \lesssim 10^7\mathrm{M}_\odot$) and star formation is widespread, LyC escapes over an extended area at the outer regions of the galaxy (in K24 this was referred to as {\it ext}-mode of LyC escape). 
In this case, the gas scale heights tend to be small ($H \lesssim 10^{21}\mathrm{cm}$), resulting in a higher gas recombination rate, with the consequence that LyC radiation is primarily absorbed by hydrogen. 
On the other hand, if most star formation takes place in highly localized areas, LyC escapes through small channels near these regions, which are usually located close to the center of the galaxy ({\it loc}-mode). In these galaxies the gas scale heights tend to be larger ($H \gtrsim 10^{21}\mathrm{cm}$), decreasing the recombination rate such that in this mode dust significantly contributes to the absorption of LyC radiation. 

\begin{figure}
    \centering
    \includegraphics[width=0.49\textwidth]{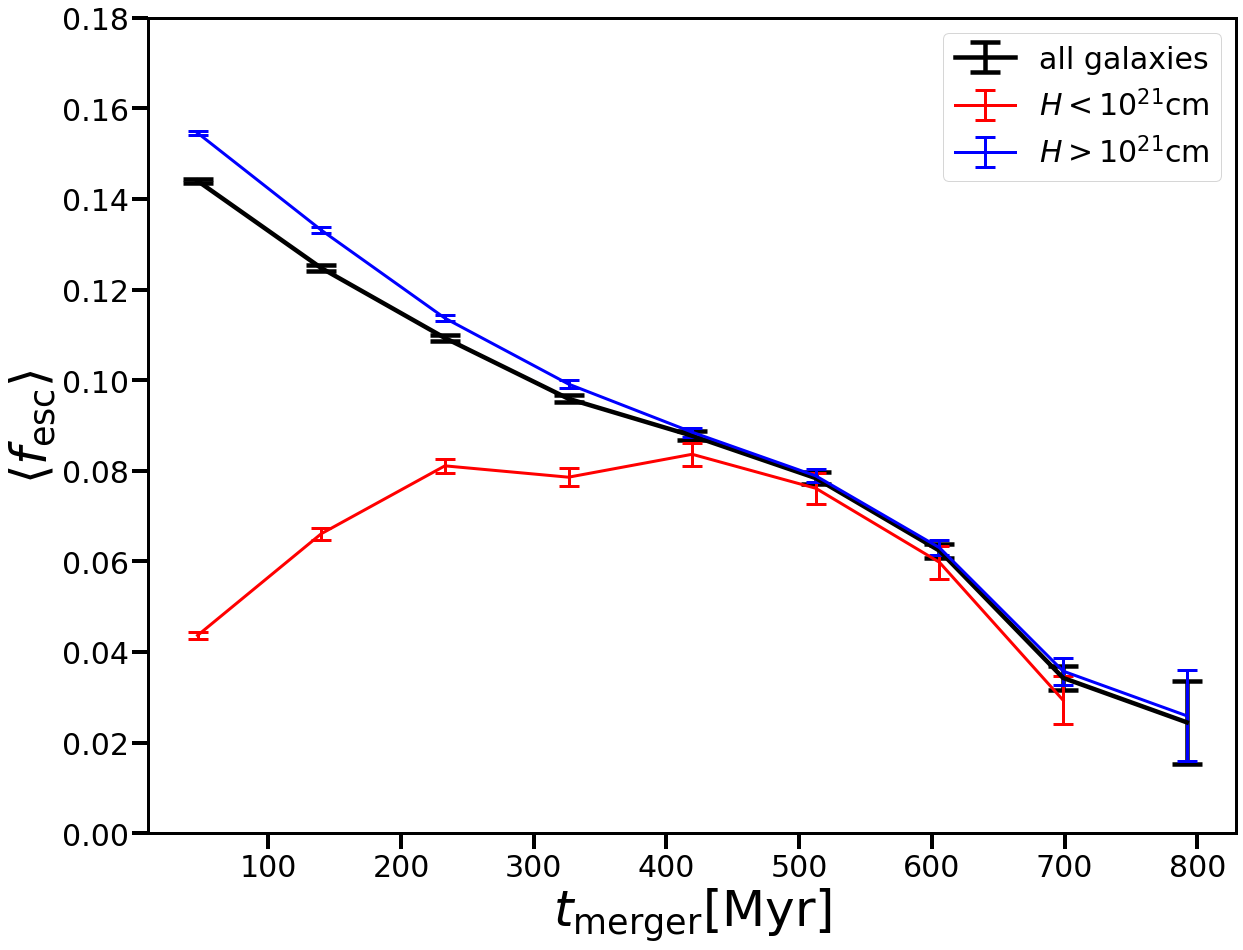}
    \caption{Average LyC escape fraction as a function of time elapsed since the last major merger for galaxies with gas scale height $H<10^{21}\mathrm{cm}$ ({\it ext}-mode of escape) and $H>10^{21}\mathrm{cm}$ ({\it loc}-mode), as well as the full sample. The lines refer to galaxies at all redshift and stellar masses.}
    \label{fig:differences_localized_extended}
\end{figure}

Due to these significant differences in the mode of LyC escape, in fig.~\ref{fig:differences_localized_extended} we investigate the correlation between the average LyC escape fraction and $t_\mathrm{merger}$ for galaxies with $H<10^{21}\mathrm{cm}$ and $H>10^{21}\mathrm{cm}$ separately. Indeed, the two modes show a very different evolution. 
As most galaxies have $H>10^{21}\mathrm{cm}$, the evolution for galaxies in which the {\it loc}-mode is prevalent looks similar to that of the full sample, as seen from the black lines in figs.~\ref{fig:redshift} and \ref{fig:mass}. 
On the other hand, for galaxies with an {\it ext}-mode of escape the correlation differs substantially at $t_\mathrm{merger} \lesssim 400\mathrm{Myr}$. 

\begin{figure}
    \centering
    \includegraphics[width=0.45\textwidth]{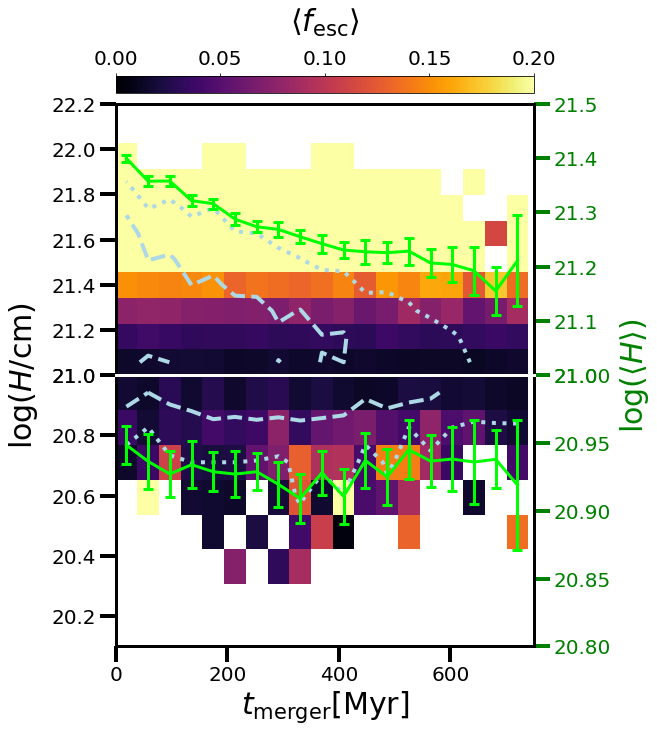}
    \caption{Histogram of the average escape fraction as a function of the time since the last major merger and the gas scale height for galaxies with $H> 10^{21}$cm ({\it loc}-mode; upper panel) and $H < 10^{21}$cm ({\it ext}-mode; lower panel). The dashed and dotted lines refer respectively to the one and two sigma distribution of galaxy counts of the respective samples, while the green line corresponds to the average gas scale height of all galaxies in a given $t_\mathrm{merger}$ bin.}
    \label{fig:column_height_modes}
\end{figure}

In fig.~\ref{fig:column_height_modes} we show histograms of the average escape fraction as a function of the time since the last major merger and the gas scale height.\footnote{The color scale is capped at $\langle f_\mathrm{esc} \rangle=0.2$ to emphasize variations in the average escape fraction. This is a purely arbitrary choice and does not imply any physical limit on $\langle f_\mathrm{esc} \rangle$} To avoid the bias introduced by the correlation between galactic stellar mass and merger time observed in fig.~\ref{fig:mass}, we examine only galaxies with  $10^{6.8}\mathrm{M}_\odot <M_\star<10^{7.2}\mathrm{M}_\odot$.
We find that the gas scale height is not greatly affected by a merger in galaxies with an {\it ext}-mode of LyC escape (upper panel), while for galaxies with {\it loc}-mode (lower) the average gas scale height immediately after a merger is higher by a factor of $1.5$ (0.2 dex) as compared to the one 400Myr after the merger. This is likely caused by tidal disruption of galactic gas during the merger, which enhances LyC escape because the recombination rate scales as $\propto H^{-2}$. 
In the \textit{ext}-mode we see that the average escape fraction is actually lower for smaller gas scale heights, as the LyC production is driven by efficient cooling on the outskirts of the galaxy as described in K24.

This example illustrates the reason for the different correlation between mergers and LyC escape observed in fig.~\ref{fig:differences_localized_extended}. 
While typically properties of galaxies exhibiting an {\it ext}-mode or {\it loc}-mode of escape are mostly affected very similarly by mergers, the different evolution of $\langle f_\mathrm{esc} \rangle$  arises from its opposite correlation with galactic properties, as illustrated in fig.~\ref{fig:column_height_modes} for the column height.

\subsection{Impact of post-merger galactic properties on the escape fraction}



The gas flow around a galaxy following a merger can have a significant impact on its star-formation activity and gas distribution, which in turn affects its LyC escape. To investigate this relationship, we define the gas flow as
\begin{equation}
    \mathcal{F_{\rm gas}} = \frac{\sum_i \vec{p}_i\cdot\vec{r}_{0,i}}{\sum_i \lvert \vec{p}_i \rvert},
\end{equation}
with $\vec{p}_i$ being the momentum of the gas particle and $\vec{r}_{0,i}$ the unit vector pointing from the galactic center to the center of mass of the gas particles. In other words, this is the ratio between the total gas momentum and the gas momentum of the flow moving away from the galaxy. Here, negative values correspond to gas inflow, with -1 meaning radial collapse, while a value of 1 indicates that all gas flows radially away from the galaxy.

\begin{figure*}
    \centering
    \includegraphics[width=\textwidth]{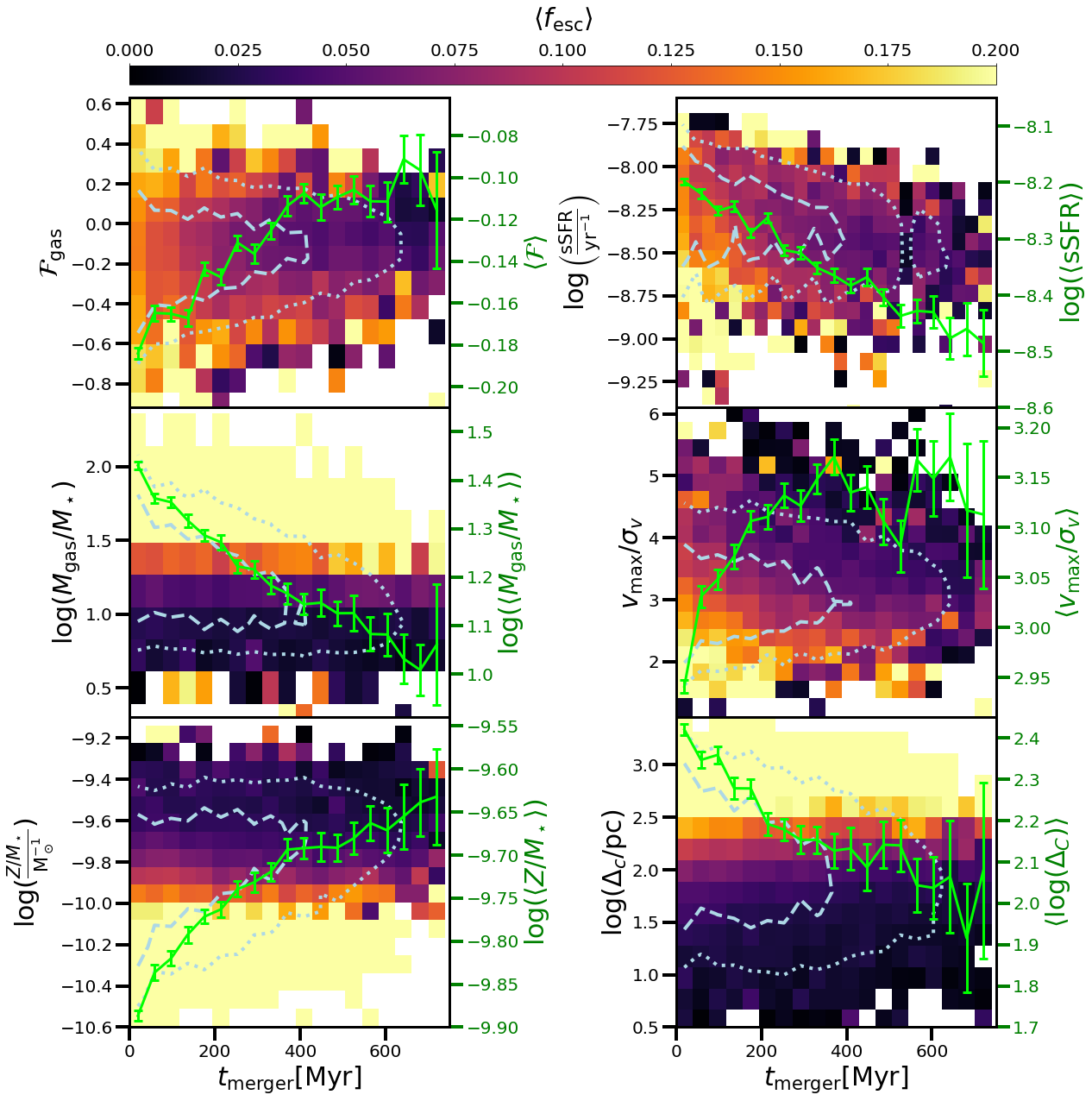}
    \caption{Histograms showing the average escape fraction (color bar) as a function of the time since the last major merger for galaxies in the \textit{loc}-mode with stellar masses $M_\star = (10^{6.8}-10^{7.2})\mathrm{M}_\odot$. The panels refer to dependance on the gas flow $\mathcal{F}_\mathrm{gas}$ (top-left), the specific SFR (top-right), the ratio of galactic gas and stellar mass (center-left), the ratio of the peak rotational velocity and the one dimensional velocity distribution (center-right), the metallicity of the galaxy normalized by the stellar mass (bottom-left) and the distance between the center of mass of the galactic gas and the center of mass of the star forming gas (bottom-right). The dashed and dotted lines refer respectively to the one and two sigma distribution of galaxy counts, while the green line represents the average value of the given quantity of all galaxies in a given $t_\mathrm{merger}$ bin.
    }
    \label{fig:gal_prop}
\end{figure*}

From the top-left panel of fig.~\ref{fig:gal_prop} we see that shortly after a merger and until $t_{\rm merger} \sim 300$~Myr, on average $\mathcal{F_{\rm gas}}$ is negative, i.e. we have predominantly gas inflow. The escape fraction has a minimum at $\mathcal{F_{\rm gas}}\sim 0$, with the value increasing for both in- and outflow. While an outflow results in lower column density and optical depth, inflow fuels star formation and the production of LyC photons, which in the \textit{loc}-mode are able to ionize escape channels. 
However, as time progresses following a merger, galaxies reach an equilibrium state. As a consequence, the gas inflow decreases, and as a result too few young stars form to ionize escape channels and the escape fraction decreases. This is likely also the reason for the flattening of the specific star formation rate (sSFR) evolution after a similar time as seen on the left. 

Next we find that on average the sSFR decreases approximately linearly on a logarithmic scale until 600~Myr after the merger, and subsequently becomes roughly constant. Indeed, directly after a merger it is twice as high ($0.3$dex) as at $t_\mathrm{merger}=600$~Myr, likely because of the additional inflow of gas onto the galactic center. 
We observe that both high and low values of sSFR lead to an elevated escape fraction. 
Notably, the location of the minimum in $\langle f_\mathrm{esc} \rangle$ as a function of sSFR decreases from $\mathrm{sSFR}=10^{-8}\mathrm{yr}^{-1}$ at $t_\mathrm{merger}=0$ Myr to $\mathrm{sSFR}=10^{-8.5}\mathrm{yr}^{-1}$ at $t_\mathrm{merger}=500$ Myr. This suggests a non-linear relationship between $f_\mathrm{esc}$ and inflow rates, which, when substantial enough to support significant star formation, results in higher escape fractions. However, when the sSFR is too low, additional inflow, while boosting the star formation rate, absorbs more LyC radiation than it generates through star formation. 



Next, we observe that following a merger event, the additional inflow of gas also results in an increase in the ratio of gas to stellar mass. Within the examined time interval this ratio increases by 0.4 dex (a factor of $\sim 2.5$). This increased availability of gas is expected to fuel star formation, thereby contributing to the formation of pathways for LyC escape. Furthermore, the inflowing gas tends to be more metal poor than the galactic gas, and thus contributes little to dust absorption. Conversely, a higher value of $M_\star$ corresponds to an increased metallicity within the galaxy, and thus enhanced dust absorption and a reduction in the overall average escape fraction.

In the center-right panel we look at the ratio $v_\mathrm{max}/\sigma_v$, where $v_\mathrm{max}$ is the maximum of the rotational velocity and $\sigma_v$ is the 1D velocity dispersion. This ratio can be used as a proxy for galactic morphology, with a higher value characterizing disc-like galaxies, while lower values corresponding to a more spherical shape.
We see that on average $v_\mathrm{max}/\sigma_v$ decreases after a recent merger, as a consequence of the tidal disruption that galaxies experience. Additionally, as discussed in the first panel, after a merger event gas flow increases, resulting in a larger velocity dispersion. 
As a consequence of the gas becoming less concentrated, its density and optical depth decrease, and thus the escape fraction increases.

In the bottom-left panel we explore the behaviour of the $t_\mathrm{merger}$ dependent escape fraction in relation to the metallicity. To remove the bias due to the correlation between $Z$ and $M_\star$, we introduce the specific metallicity as the metallicity normalized by the stellar mass ($Z/M_\star$). This quantity decreases on average by approximately 0.3dex (equivalent to a factor of 2) within the examined $t_\mathrm{merger}$ range. This decrease in the specific metallicity is particularly pronounced in the first 300~Myr after a merger. This in turn results in an increase in the average escape fraction of galaxies as can be seen by the contour lines encompassing a larger number of the high $\langle f_\mathrm{esc} \rangle$ bins. The cause for this correlation lies in the previously mentioned migration of metal-depleted gas from the CGM toward the inner galactic regions during the merger process. As discussed in K24, dust in the {\it loc-}mode significantly contributes to LyC absorption for galaxies in the examined mass range. Given our assumption of a linear relation between metal and dust content, a negative correlation emerges between the specific metallicity and LyC escape. 

Finally, \cite{lereste2023} have suggested that mergers result in an offset of gas from the star-forming regions, thus facilitating the formation of ionized channels and the escape of radiation.
To investigate this, we introduce the quantity $\Delta_C$, which is defined as the distance between the center of mass of all gas within the galaxy and the center of mass of the star-forming gas.
Indeed, in the lower-right panel, we see that $\Delta_C$  increases after a merger, and that a large offset is strongly correlated with an increase in the escape fraction, which is to be expected because of the reduction in optical depth due to the lower column density. We therefore conclude that for small $t_\mathrm{merger}$, neutral gas is initially offset from the central star forming regions, but momentum exchange between neutral and star-forming gas slowly reduces this offset leading to a lower $\Delta_C$.

\subsection{Mergers and increased Ly$\alpha$ visibility}

\begin{figure}
    \centering
    \includegraphics[width=0.5\textwidth]{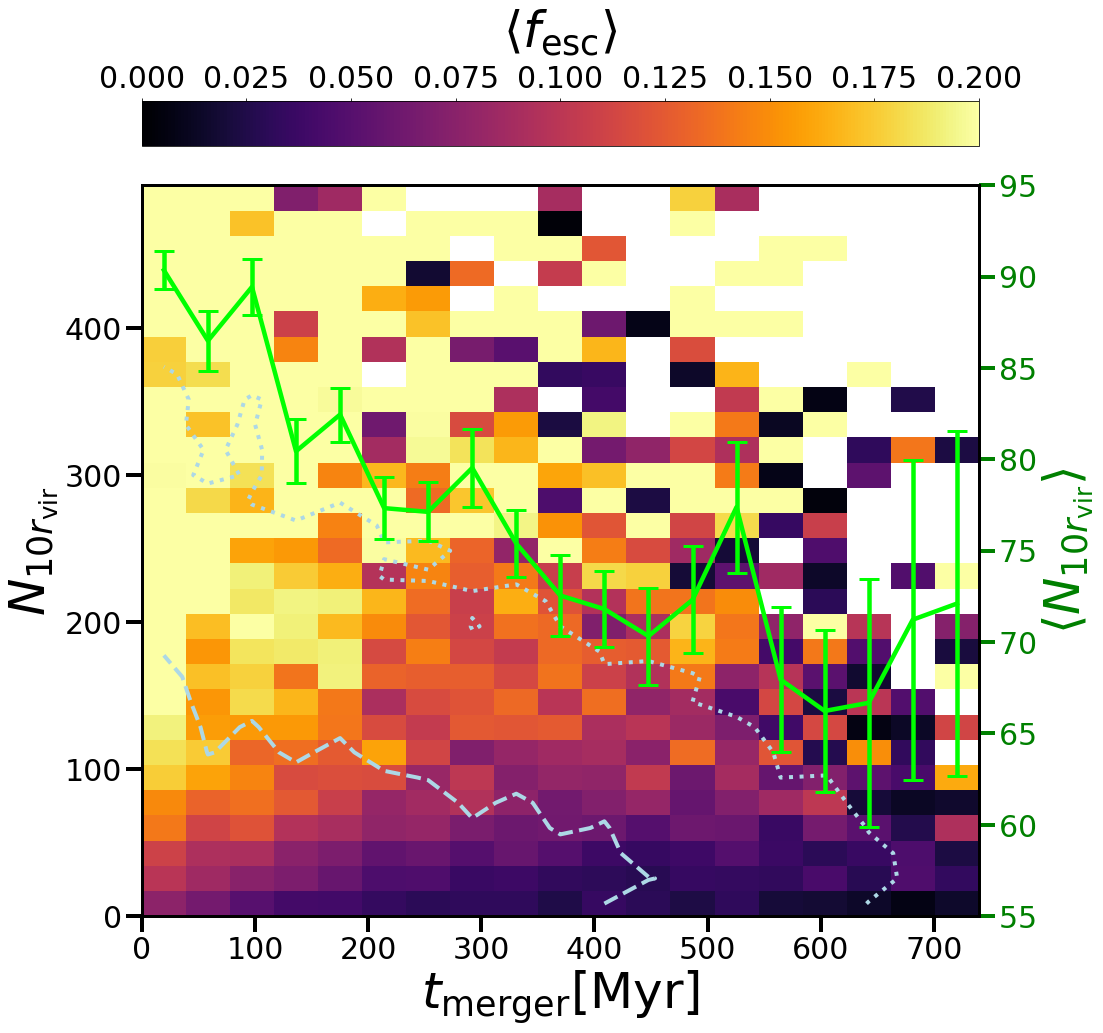}
    \caption{Number of galactic neighbors within 10 virial radii of a galaxy as a function of $t_\mathrm{merger}$ in the \textit{loc}-mode. 
    The dashed and dotted lines refer respectively to the one and two sigma distribution of galaxy counts. The green line corresponds to the average value of $N_{10r_\mathrm{vir}}$ of all galaxies in a given $t_\mathrm{merger}$ bin. We inspect galaxies at all redshift with $M_\star = (10^{6.8}-10^{7.2})\mathrm{M}_\odot$.
    }
    \label{fig:proximity}
\end{figure}

Finally, we test the suggestion made in \cite{witten23} that a merger driven increase in LyC escape might facilitate the creation of ionized proximity zones surrounding high-$z$ Ly$\alpha$ emitters (LAEs), increasing their visibility. To investigate this hypothesis, in Fig.~\ref{fig:proximity} we show how $\langle f_{\rm esc}\rangle$ and $t_{\rm merger}$ are correlated to the average number of neighbours a galaxy possesses within 10 virial radii $N_{10r_\mathrm{vir}}$\footnote{As the virial radius is not provided for TNG subhalos, we used the stellar half mass radius as a proxy, assuming $r_\mathrm{SHMR}=0.015r_\mathrm{vir}$\citep{kravtsov13}.}, which is a proxy to determine whether a galaxy is located in an over- or underdense region. To ensure that only neighbors that would result in a significant merger are taken into account, we consider only galaxies with masses that are within $0.4$dex below and $0.2$dex above\footnote{The asymmetric binning is based on the idea that the galaxy emerged through the merger of two lighter galaxies. Therefore, slightly smaller galaxies are a better tracer of the merger frequency experienced by a galaxy in a given mass bin.}
the mass of the galaxy under consideration. As before, we perform this analysis for galaxies with $M_\star =  10^{6.8}-10^{7.2}$~M$_\odot$. Finally, as we found no significant qualitative variation with redshift, we increase the statistics by considering all redshifts together.

The average number of neighbors decreases for larger $t_\mathrm{merger}$, from 90 for galaxies right after a merger to 65 at $t_\mathrm{merger}=600$~Myr, supporting the expectation that mergers take place more frequently in overdense environments. The color gradient along the x-axis indicates that for a fixed number of neighbors, recent mergers increase the escape fraction, supporting the suggestion by \cite{witten23} that the more frequent mergers in overdense environments contribute to an increase in LyC escape. 
However, we additionally see a strong color gradient along the y-axis, indicating that independently from the last merger, galaxies with more neighbors tend to have higher LyC escape. We have verified that this is likely to be caused by more efficient gas exchange with the environment for galaxies in overdense regions, resulting in higher gas to star ratios as well as lower metallicities, which are both correlated with higher escape fractions.

\section{Conclusions}
\label{sec:concl}

Recent observational studies (e.g. \citealt{lereste2023} and \citealt{witten23}) have suggested that mergers of dwarf galaxies at high redshift might be a key cause of LyC leakage. The latter would be fueled by the increase in star formation associated  to gas in-flow, the reduction of gas column density due to tidal disruptions, as well as the offset of neutral gas from the star-forming regions. 
In this study we have applied the method introduced in Kostyuk et al. (2023b) to TNG50 galaxies, and confirmed that also in these simulations LyC escape increases following recent merger events.

We find that the effect is most pronounced in galaxies with $M_\star \approx 10^7\mathrm{M}_\odot$, for which the average escape fraction decreases from 14\% immediately after a merger to 3\% 700~Myr after the event. It is worth noting that mergers may exert a significant influence even on smaller galaxies, but the effect is expected to occur over timescales shorter than those resolved by the simulation. On the other hand, the absolute decrease in the escape fraction is smaller for higher mass galaxies, given their intrinsically lower escape fraction.

While mergers affect the physical properties of all galaxies in a similar way, the change of these properties has a different impact on the LyC escape, depending on whether a galaxy belongs to the \textit{ext}- or \textit{loc}-mode.
Indeed, we find that galactic mergers lead to an increase in $f_{\rm esc}$ only in the more prevalent {\it loc}-mode, where LyC leakage takes place mainly from relatively small star-forming regions in the center of the galaxy. 
In the {\it ext}-mode, instead, where LyC leaks from wide star-forming regions in the outskirts of the galaxy, we found that a merger event decreases the escape fraction in the \textit{ext}-mode. In this mode the escape fraction is higher $\sim 200$~Myr following a merger as compared to immediately after a merger event.

To gain deeper insights into the relationship between LyC escape and galactic mergers, we investigated the correlation between post-merger galactic properties and the resulting LyC escape. We have found that, following a merger, metal poor gas flows from the galactic environment into the galactic center, increasing star formation and thus LyC production, which  in turn creates more escape channels. Additionally, mergers cause a displacement of neutral gas relative to the star-forming gas, which in turn reduces the opacity for radiation escaping the galaxy.

Finally, we introduced the average number of neighbors a galaxy has within 10 virial radii as a measure for the galactic density of the environment. We found that galaxies having more neighbors tend to have higher escape fractions, due to the more frequent mergers. Additionally, LyC escape from galaxies in overdense regions is increased because of the larger rate of gas inflow, which increases the star-formation rate while decreasing gas absorption due to decreasing metallicity.

In summary, our findings support the observational claims that in the early Universe, merging of dwarf galaxies leads to a substantial enhancement in the escape of LyC radiation.

In the future, we aim to enhance the robustness of our findings by extending our analysis to a wider range of high redshift simulations in order to investigate the model dependency of our results.

\section*{Data Availability}
The data that support the findings of this study are available upon request from the authors.

\begin{acknowledgements}
We thank R{\"u}diger Pakmor, Aniket Bhagwat, Tiago Costa and Enrico Garaldi for insightful discussions and valuable feedback.
\end{acknowledgements}

\bibliographystyle{aa}
\bibliography{refs}

\end{document}